\documentclass[12pt]{article}%
\usepackage{amsmath}
\usepackage{amsfonts}
\usepackage{amssymb}
\usepackage{graphicx}%
\newtheorem{theorem}{Theorem}

\newtheorem{definition}[theorem]{Definition}

\begin{document}

\title{On the solution of the static Maxwell system in axially symmetric
inhomogeneous media}
\author{Kira V. Khmelnytskaya$^{\text{1}}$, Vladislav V. Kravchenko$^{\text{2}}$,
H\'{e}ctor Oviedo$^{\text{3}}$\\$^{\text{1}}${\small CINVESTAV-Queretaro, Libramiento Norponiente No. 2000, }\\{\small Fracc. Real de Juriquilla, Queretaro, Qro. C.P. 76230 MEXICO}\\$^{\text{2}}${\small Department of Mathematics, CINVESTAV-Queretaro, }\\{\small Libramiento Norponiente No. 2000, Fracc. Real de Juriquilla,
Queretaro, }\\{\small Qro. C.P. 76230 MEXICO e-mail: vkravchenko@qro.cinvestav.mx}\\$^{\text{3}}${\small SEPI, ESIME Zacatenco, Instituto Polit\'{e}cnico
Nacional, Av. IPN S/N, }\\{\small C.P. 07738, D.F. MEXICO\thanks{Research was supported by CONACYT,
Mexico }}}
\maketitle

\begin{abstract}
We consider the static Maxwell system with an axially symmetric dielectric
permittivity and construct complete systems of its solutions which can be used
for analytic and numerical solution of corresponding boundary value problems.

\end{abstract}

\section{Introduction}

Consider the static Maxwell system%
\begin{equation}
\operatorname*{div}(\varepsilon\mathbf{E})=0,\qquad\operatorname*{rot}%
\mathbf{E}=0 \label{MaxwellStatic}%
\end{equation}
where we suppose that $\varepsilon$ is a function of the cylindrical radial
variable $r=\sqrt{x_{1}^{2}+x_{2}^{2}}$: $\varepsilon=\varepsilon(r)$. Two
important situations are usually studied: the meridional field and the
transverse field.

The first case is characterized by the condition that the vector $\mathbf{E}$
is independent of the angular coordinate $\theta$ and the component
$E_{\theta}$ of the vector $\mathbf{E}$ in cylindrical coordinates vanishes
identically. The vector of such field belongs to a plane containing the axis
$x_{3}$ and depends only on the distance $r$ to this axis as well as on the
coordinate $x_{3}$. The field then is completely described by a two-component
vector-function in the plane $(r,x_{3})$.

The second case is characterized by the condition that the vector $\mathbf{E}$
is independent of $x_{3}$ and the component $E_{3}$ is identically zero. The
vector of such field belongs to a plane perpendicular to the axis $x_{3}$ and
the corresponding model reduces to a two-component vector-function in the
plane $(x_{1},x_{2})$.

In the present work in both cases we construct a complete system of solutions
of the corresponding model. We use the fact that in both cases the system
(\ref{MaxwellStatic}) reduces to a system describing so-called $p$-analytic
functions \cite{AS}, \cite{Chemeris}, \cite{Goman}, \cite{KapshKlen},
\cite{KapshYazk}, \cite{KrPanalyt}, \cite{Polozhy},
\cite{ZabarankinUlitko2006}
\begin{equation}
u_{x}=\frac{1}{p}v_{y},\qquad u_{y}=-\frac{1}{p}v_{x}.\qquad\label{Polozh}%
\end{equation}
In the first case the function $p$ is a function of one Cartesian variable $x$
meanwhile in the second it is a function of $r=\sqrt{x_{{}}^{2}+y_{{}}^{2}}$.
In both cases we construct an infinite system of so-called formal powers
\cite{Berskniga}, \cite{Courant}. This is a complete system of exact solutions
of equations (\ref{Polozh}) generalizing the system of usual complex powers
$(z-z_{0})^{n},$ $n=0,1,2,\ldots$. Locally, near the center $z_{0}$ the formal
powers behave asymptotically like powers. Nevertheless in general their
behaviour can be arbitrarily different from that of powers but with a
guarantee of their completeness in the sense that any solution of the
considered equations can be represented as a uniformly convergent series of
formal powers. The general theory of formal powers was developed by L. Bers
\cite{Berskniga} as a part of his pseudoanalytic function theory. Its
application was restricted by the fact that only for a quite limited class of
pseudoanalytic functions the explicit construction of formal powers was
possible. L. Bers' results allow us to construct a complete system of formal
powers in the meridional case. Nevertheless they are not applicable to the
model arising from the transverse case. In the recent works \cite{Krpseudoan}
and \cite{KrRecentDevelopments} the class of solvable in this sense systems
(\ref{Polozh}) was substantially extended. In the present work we use these
results for solving the static Maxwell system in the transverse case. This
combination of the relation between the static Maxwell system
(\ref{MaxwellStatic}) and the system (\ref{Polozh}) together with the
classical results of L. Bers on pseudoanalytic formal powers and new
developments in \cite{Krpseudoan} and \cite{KrRecentDevelopments} allow us to
obtain a general solution of the static Maxwell system in the axially
symmetric case in the sense that we construct a complete system of its
solutions for both the meridional and the transverse fields.

\section{Reduction of the static Maxwell system to $p$-analytic functions}

\subsection{The meridional case}

Introducing the cylindrical coordinates and making the assumptions that
$\mathbf{E}$ is independent of the angular variable $\theta$ and that the
component $E_{\theta}$ is identically zero we obtain that (\ref{MaxwellStatic}%
) can be written as follows%
\[
\frac{\partial E_{r}}{\partial x_{3}}-\frac{\partial E_{3}}{\partial
r}=0,\qquad\frac{1}{r}\frac{\partial(r\varepsilon E_{r})}{\partial r}%
+\frac{\partial(\varepsilon E_{3})}{\partial x_{3}}=0.
\]
Denote $x=r$, $y=x_{3}$, $u=E_{3}$ and $v=r\varepsilon E_{r}$. Then the system
takes the form%
\[
u_{x}=\frac{1}{x\varepsilon(x)}v_{y},\qquad u_{y}=-\frac{1}{x\varepsilon
(x)}v_{x},
\]
where the subindices denote the derivatives with respect to the corresponding
variables. Thus, in the case of a meridional field the vector $\mathbf{E}$ is
completely described by an $x\varepsilon(x)$-analytic function $\omega=u+iv$.

\subsection{The transverse case}

We assume that $\mathbf{E}$ is independent of the longitudinal variable
$x_{3}$ and $E_{3}\equiv0$. Then from (\ref{MaxwellStatic}) we have that the
vector $(E_{1},E_{2})^{T}$ is the gradient of a function $u=u(x_{1},x_{2})$
which satisfies the two-dimensional equation%
\begin{equation}
\operatorname*{div}(\varepsilon\nabla u)=0. \label{condeps}%
\end{equation}
Denote $x=x_{1}$, $y=x_{2}$, $z=x+iy$ and consider the system%
\begin{equation}
u_{x}=\frac{1}{\varepsilon}v_{y},\qquad u_{y}=-\frac{1}{\varepsilon}v_{x}.
\label{p=eps}%
\end{equation}
It is easy to see that if the function $\omega=u+iv$ is its solution then $u$
is a solution of (\ref{condeps}), and vice versa \cite{KrJPhys06}, if $u$ is a
solution of (\ref{condeps}) in a simply connected domain $\Omega$ then
choosing
\begin{equation}
v=\overline{A}(i\varepsilon u_{\overline{z}}), \label{reconstv}%
\end{equation}
where
\begin{equation}
\overline{A}\left[  \Phi\right]  (x,y)=2\left(  \int_{\Gamma}%
\operatorname*{Re}\Phi dx+\operatorname*{Im}\Phi dy\right)  +c, \label{APhi}%
\end{equation}
$c$ is an arbitrary real constant, $\Gamma$ is an arbitrary rectifiable curve
in $\Omega$ leading from $(x_{0},y_{0})$ to $(x,y)$ we obtain that
$\omega=u+iv$ is a solution of (\ref{p=eps}). Here the subindex $\overline{z}$
means the application of the operator $\partial_{\overline{z}}=\frac{1}%
{2}(\partial_{x}+i\partial_{y})$. Note that due to the fact that $u$ is a
solution of (\ref{condeps}) the function $\Phi=i\varepsilon u_{\overline{z}}$
satisfies the condition $\partial_{y}\operatorname*{Re}\Phi-\partial
_{x}\operatorname*{Im}\Phi=0$ and hence the integral in (\ref{APhi}) is
path-independent. For a convex domain then expression (\ref{APhi}) can be
written as follows%
\[
\overline{A}\left[  \Phi\right]  (x,y)=2\left(  \int_{x_{0}}^{x}%
\operatorname*{Re}\Phi(\eta,y)d\eta+\int_{y_{0}}^{y}\operatorname*{Im}%
\Phi(x_{0},\xi)d\xi\right)  +c.
\]
Note that $v$ is a solution of the equation
\[
\operatorname*{div}(\frac{1}{\varepsilon}\nabla v)=0.
\]
Thus, equation (\ref{condeps}) (and hence the system (\ref{MaxwellStatic}) in
the case under consideration) is equivalent to the system (\ref{p=eps}) in the
sense that if $\omega=u+iv$ is a solution of (\ref{p=eps}) then its real part
$u$ is a solution of (\ref{condeps}) and vice versa, if $u$ is a solution of
(\ref{condeps}) then $\omega=u+iv,$ where $v$ is constructed according to
(\ref{reconstv}) is a solution of (\ref{p=eps}).

We reduced both considered cases the meridional and the transverse to the
system describing $p$-analytic functions. In the first case $p=x\varepsilon
(x)$ is a function of one Cartesian variable and in the second $p=\varepsilon
(r)$, $r=\sqrt{x_{{}}^{2}+y_{{}}^{2}}$. As we show below in both cases we are
able to construct explicitly a complete system of formal powers and hence a
complete system of exact solutions of the corresponding Maxwell system. Let us
notice that equation (\ref{condeps}) with $\varepsilon$ being a function of
the variable $r$ was considered in the recent work \cite{Demidenko} whith
applications to electrical impedance tomography. The algorithm proposed in
that work implies numerical solution of a number of ordinary differential
equations arising after a standard separation of variables. Our construction
of a complete system of solutions of (\ref{condeps}) is based on essentially
different ideas and does not require solving numerically any differential equation.

\section{$p$-analytic functions and formal powers}

\subsection{The main Vekua equation\label{SubsectMainVekua}}

Consider the system describing $p$-analytic functions%
\begin{equation}
u_{x}=\frac{1}{p}v_{y},\qquad u_{y}=-\frac{1}{p}v_{x}, \label{p-anal}%
\end{equation}
where we suppose that $p$ is a positive and continuously differentiable
function of $x$ and $y$. Together with tis system we consider the following
Vekua equation which due to its importance in relation to second-order
elliptic equations of mathematical physics is called \cite{KrJPhys06},
\cite{KrRecentDevelopments} the main Vekua equation
\begin{equation}
W_{\overline{z}}=\frac{f_{\overline{z}}}{f}\overline{W}, \label{Vekuamain1}%
\end{equation}
where $f=\sqrt{p}$. The function $\omega=u+iv$ is a solution of (\ref{p-anal})
iff \cite{KrRecentDevelopments} $W=uf+iv/f$ is a solution of (\ref{Vekuamain1}).

In \cite{KrRecentDevelopments} there was proposed a method for explicit
construction of the system of formal powers corresponding to the main Vekua
equation under a quite general condition on $f$. Here we briefly describe the
method for which we need first to recall some basic definitions from L. Bers'
theory of formal powers \cite{Berskniga}.

Let $F$ and $G$ be a couple of solutions of a Vekua equation
\begin{equation}
W_{\overline{z}}=a_{(F,G)}W+b_{(F,G)}\overline{W}\text{\qquad in }%
\Omega\label{VekuaGen}%
\end{equation}
such that $\operatorname{Im}(\overline{F}G)>0.$ Then $(F,G)$ is said to be a
\textit{generating pair} corresponding to (\ref{VekuaGen}). The complex
functions $a_{(F,G)}$ and $b_{(F,G)}$ are called \textit{characteristic
coefficients }of the pair $(F,G)$ and it can be seen that
\[
a_{(F,G)}=-\frac{\overline{F}G_{\overline{z}}-F_{\overline{z}}\overline{G}%
}{F\overline{G}-\overline{F}G},\qquad b_{(F,G)}=\frac{FG_{\overline{z}%
}-F_{\overline{z}}G}{F\overline{G}-\overline{F}G}.
\]
Together with these characteristic coefficients another pair of characteristic
coefficients is introduced in relation to the notion of the $(F,G)$-derivative:%

\[
A_{(F,G)}=-\frac{\overline{F}G_{z}-F_{z}\overline{G}}{F\overline{G}%
-\overline{F}G},\qquad B_{(F,G)}=\frac{FG_{z}-F_{z}G}{F\overline{G}%
-\overline{F}G},
\]
where the $z$ means the application of the operator $\partial_{z}=\frac{1}%
{2}\left(  \frac{\partial}{\partial x}-i\frac{\partial}{\partial y}\right)  $.
As in the present work we do not use explicitly the notion of the
$(F,G)$-derivative, we refer the interested reader to \cite{Berskniga} for its
definition and properties. However we do need the concept of characteristic
coefficients for defining the following important object.

\begin{definition}
\label{DefSuccessor}Let $(F,G)$ and $(F_{1},G_{1})$ - be two generating pairs
in $\Omega$ corresponding to the Vekua equations with coefficients $a_{(F,G)}%
$, $b_{(F,G)}$ and $a_{(F_{1},G_{1})}$ and $b_{(F_{1},G_{1})}$ respectively.
Then $(F_{1},G_{1})$ is called \ successor of $(F,G)$ and $(F,G)$ is called
predecessor of $(F_{1},G_{1})$ if%
\[
a_{(F_{1},G_{1})}=a_{(F,G)}\qquad\text{and}\qquad b_{(F_{1},G_{1})}%
=-B_{(F,G)}\text{.}%
\]

\end{definition}

\begin{definition}
\label{DefSeq}A sequence of generating pairs $\left\{  (F_{m},G_{m})\right\}
$, $m=0,\pm1,\pm2,\ldots$ , is called a generating sequence if $(F_{m+1}%
,G_{m+1})$ is a successor of $(F_{m},G_{m})$. If $(F_{0},G_{0})=(F,G)$, we say
that $(F,G)$ is embedded in $\left\{  (F_{m},G_{m})\right\}  $.
\end{definition}

For any generating pair $(F,G)$ the corresponding $(F,G)$-integral is defined
as follows%
\[
\int_{\Gamma}wd_{(F,G)}\zeta=F(z)\operatorname{Re}\int_{\Gamma}G^{\ast}%
wd\zeta+G(z)\operatorname{Re}\int_{\Gamma}F^{\ast}wd\zeta
\]
where $\Gamma$ is a rectifiable curve leading from $z_{0}$ to $z$ and
$(F^{\ast},G^{\ast})$ is an adjoint generating pair defined by the equations%
\[
F^{\ast}=-\frac{2\overline{F}}{F\overline{G}-\overline{F}G},\qquad G^{\ast
}=\frac{2\overline{G}}{F\overline{G}-\overline{F}G}.
\]
If $w$ is an $(F_{1},G_{1})$ - pseudoanalytic function (i.e., it is a solution
of the Vekua equation with the coefficients $a_{(F_{1},G_{1})}$ and
$b_{(F_{1},G_{1})}$) then its $(F,G)$-integral is path-independent.

Now we are ready to introduce the definition of formal powers.

\begin{definition}
\label{DefFormalPower}The formal power $Z_{m}^{(0)}(a,z_{0};z)$ with center at
$z_{0}\in\Omega$, coefficient $a$ and exponent $0$ is defined as the linear
combination of the generators $F_{m}$, $G_{m}$ with real constant coefficients
$\lambda$, $\mu$ chosen so that $\lambda F_{m}(z_{0})+\mu G_{m}(z_{0})=a$. The
formal powers with exponents $n=1,2,\ldots$ are defined by the recursion
formula%
\begin{equation}
Z_{m}^{(n)}(a,z_{0};z)=n\int_{z_{0}}^{z}Z_{m+1}^{(n-1)}(a,z_{0};\zeta
)d_{(F_{m},G_{m})}\zeta. \label{recformula}%
\end{equation}

\end{definition}

This definition implies the following properties.

\begin{enumerate}
\item $Z_{m}^{(n)}(a,z_{0};z)$ is an $(F_{m},G_{m})$-pseudoanalytic function
of $z$.

\item If $a^{\prime}$ and $a^{\prime\prime}$ are real constants, then
$Z_{m}^{(n)}(a^{\prime}+ia^{\prime\prime},z_{0};z)=a^{\prime}Z_{m}%
^{(n)}(1,z_{0};z)+a^{\prime\prime}Z_{m}^{(n)}(i,z_{0};z).$

\item The asymptotic formulas
\begin{equation}
Z_{m}^{(n)}(a,z_{0};z)\sim a(z-z_{0})^{n},\quad z\rightarrow z_{0}
\label{asymptformulas}%
\end{equation}
hold.
\end{enumerate}

Writing $Z_{{}}^{(n)}(a,z_{0};z)$ we indicate that the formal power
corresponds to the generating pair $(F,G)$.

The definition of formal powers shows us that in order to obtain $Z_{{}}%
^{(n)}(a,z_{0};z)$ we need to have first the formal power $Z_{1}%
^{(n-1)}(a,z_{0};z)$ for which it is necessary to calculate $Z_{2}%
^{(n-2)}(a,z_{0};z)$ and so on. Thus, the problem of construction of formal
powers of any order for a given generating pair $(F,G)$ reduces to the
construction of a corresponding generating sequence. Then definition
\ref{DefFormalPower} gives us a simple algorithm for constructing the formal
powers. In other words, one needs a pair of exact solutions for each of the
infinite number of Vekua equations corresponding to a generating sequence.

In the next subsection we show how this seemingly difficult task can be
accomplished in a quite general situation. Meanwhile here we recall some well
known results in order to explain that the system of formal powers in fact
represents a complete system of solutions of a corresponding Vekua equation.
First of all, let us notice that due to the property 2 of formal powers for
every $n$ (and for a fixed $z_{0}$) it is sufficient to construct only two
formal powers: $Z_{{}}^{(n)}(1,z_{0};z)$ and $Z_{{}}^{(n)}(i,z_{0};z)$, then
for any coefficient $a$ the corresponding formal power $Z_{{}}^{(n)}%
(a,z_{0};z)$ is a linear combination of the former two.

An expression of the form $\sum_{n=0}^{N}Z_{{}}^{(n)}(a_{n},z_{0};z)$ is
called a \textit{formal polynomial. }Under the conditions imposed in this work
on the function $\varepsilon$ and on the domain of interest $\Omega$ (see
section 4 and for more details \cite{KrRecentDevelopments}) the following
Runge-type theorem is valid where following \cite{Berskniga} we say that a
series converges normally in a domain $\Omega$ if it converges uniformly on
every bounded closed subdomain of $\Omega$.

\begin{theorem}
\cite{BersFormalPowers}\label{ThRunge} A pseudoanalytic function defined in a
simply connected domain can be expanded into a normally convergent series of
formal polynomials.
\end{theorem}

In other words a pseudoanalytic function can be represented as an infinite
linear combination of the functions%
\[
\left\{  Z_{{}}^{(n)}(1,z_{0};z),\quad Z_{{}}^{(n)}(i,z_{0};z)\right\}
_{n=0}^{\infty}.
\]
Moreover, if we know that a pseudoanalytic function $W$ satisfies the
H\"{o}lder condition on the boundary of a domain of interest $\Omega$ (a
common requirement when a boundary value problem is considered) then, e.g.,
the following estimate in the $C(\overline{\Omega})$-norm is available.

\begin{theorem}
\cite{Menke}\label{ThMenke} Let $W$ be a pseudoanalytic function in a domain
$\Omega$ bounded by a Jordan curve and satisfy the H\"{o}lder condition on
$\partial\Omega$ with the exponent $\alpha$ ($0<\alpha\leq1$). Then for any
$\epsilon>0$ and any natural $n$ there exists a pseudopolynomial of order $n$
satisfying the inequality
\[
\left\vert W(z)-P_{n}(z)\right\vert \leq\frac{\operatorname*{Const}}%
{n^{\alpha-\epsilon}}\qquad\text{for any }z\in\overline{\Omega}%
\]
where the constant does not depend on $n$, but only on $\epsilon$.
\end{theorem}

These and other results on interpolation and on the degree of approximation by
pseudopolynomials which can be found in the vast bibliography dedicated to
pseudoanalytic function theory (see, e.g., \cite{Fryant}, \cite{IsmTagieva})
show us that the system of formal powers is as good for solving corresponding
boundary value problems as is the system of usual complex powers
$(z-z_{0})^{n},$ $n=0,1,2,\ldots$. The real (or imaginary) parts of $\left\{
(z-z_{0})^{n}\right\}  _{n=0}^{\infty}$ are harmonic polynomials successfully
applied to the numerical solution of boundary value problems for the Laplace
equation. In a similar way the real parts of formal powers $\left\{  Z_{{}%
}^{(n)}(1,z_{0};z),\quad Z_{{}}^{(n)}(i,z_{0};z)\right\}  _{n=0}^{\infty}$
corresponding to the main Vekua equation (\ref{Vekuamain1}) (where
$f=\sqrt{\varepsilon}$) can be used for the numerical solution of boundary
value problems for the conductivity equation (\ref{condeps}) because as was
shown in \cite{KrJPhys06} the system of functions
\[
\left\{  \frac{1}{\sqrt{\varepsilon}}\operatorname*{Re}Z_{{}}^{(n)}%
(1,z_{0};z),\quad\frac{1}{\sqrt{\varepsilon}}\operatorname*{Re}Z_{{}}%
^{(n)}(i,z_{0};z)\right\}  _{n=0}^{\infty}%
\]
is complete in the space of solutions of (\ref{condeps}) in the sense of
theorems \ref{ThRunge} and \ref{ThMenke}.

A formal power $Z_{{}}^{(n)}(a,z_{0};z)$ related to the Vekua equation
(\ref{Vekuamain1}) corresponds to a formal power $_{\ast}Z_{{}}^{(n)}%
(a,z_{0};z)$ (we use the notation of L. Bers) related to the system
(\ref{p-anal}) in the following way%
\[
_{\ast}Z_{{}}^{(n)}(a,z_{0};z)=\frac{1}{f}\operatorname*{Re}Z_{{}}%
^{(n)}(a,z_{0};z)+if\operatorname*{Im}Z_{{}}^{(n)}(a,z_{0};z).
\]
As\ we will see in the meridional case it is convenient to work directly with
formal powers $_{\ast}Z_{{}}^{(n)}(a,z_{0};z)$. Any solution $\omega=u+iv$ of
the system (\ref{p-anal}) can be expanded into a normally convergent series of
real linear combinations of the complex functions $\left\{  _{\ast}Z_{{}%
}^{(n)}(1,z_{0};z),\quad_{\ast}Z_{{}}^{(n)}(i,z_{0};z)\right\}  $.

\subsection{Construction of generating sequences}

In \cite{KrRecentDevelopments} the following result was obtained.

\begin{theorem}
\label{ThGenSeq} Let $F=U(u)V(v)$ and $G=\frac{i}{U(u)V(v)}$ where $U$ and $V$
are arbitrary differentiable nonvanishing real valued functions, $\Phi=u+iv$
is an analytic function of the variable $z=x+iy$ in $\Omega$ such that
$\Phi_{z}$ is bounded and has no zeros in $\Omega$. Then the generating pair
$(F,G)$ is embedded in the generating sequence $(F_{m},G_{m})$, $m=0,\pm
1,\pm2,\ldots$in $\Omega$ defined as follows
\[
F_{m}=\left(  \Phi_{z}\right)  ^{m}F\quad\text{and\quad}G_{m}=\left(  \Phi
_{z}\right)  ^{m}G\quad\text{for even }m
\]
and%
\[
F_{m}=\frac{\left(  \Phi_{z}\right)  ^{m}}{U^{2}}F\quad\text{and\quad}%
G_{m}=\left(  \Phi_{z}\right)  ^{m}U^{2}G\quad\text{for odd }m.
\]

\end{theorem}

This theorem opens the way for construction of generating sequences and
consequently of formal powers in a quite general situation (see
\cite{KrRecentDevelopments}) and in particular in both cases considered in the
present work. In the meridional case the theorem reduces to the result of L.
Bers \cite{Berskniga} which we use in the next subsection while in the
transverse case this and other classical results are insufficient for
constructing formal powers explicitly and theorem \ref{ThGenSeq} is indispensable.

\section{Construction of formal powers}

\subsection{Formal powers in the meridional case}

As was shown in subsection 2.1 in the meridional case the Maxwell system
reduces to the following couple of equations
\[
u_{x}=\frac{1}{x\varepsilon(x)}v_{y},\qquad u_{y}=-\frac{1}{x\varepsilon
(x)}v_{x}%
\]
which is equivalent to the system considered in \cite[N18.1]{Berskniga}
\[
\sigma(x)\phi_{x}=\tau(y)\psi_{y},\qquad\sigma(x)\phi_{y}=-\tau(y)\psi_{x}.
\]
Taking $\sigma(x)=x\varepsilon(x)$ and $\tau\equiv1$ we can use the elegant
formulas for the generating powers obtained by L. Bers. Let%

\[
X^{(0)}(x_{0},x)=\widetilde{X}^{(0)}(x_{0},x)=1
\]
and for $n=1,2,...$denote%

\[
X^{(n)}(x_{0},x)=n%
{\displaystyle\int\limits_{x_{0}}^{x}}
X^{(n-1)}(x_{0},t)\frac{1}{t\varepsilon(t)}dt\text{ \ \ }\quad\text{for odd
}n
\]

\[
X^{(n)}(x_{0},x)=n%
{\displaystyle\int\limits_{x_{0}}^{x}}
X^{(n-1)}(x_{0},t)t\varepsilon(t)dt\text{ \ \ }\quad\text{for even }n
\]

\[
\widetilde{X}^{(n)}(x_{0},x)=n%
{\displaystyle\int\limits_{x_{0}}^{x}}
\widetilde{X}^{(n-1)}(x_{0},t)t\varepsilon(t)dt\text{ \ \ }\quad\text{for odd
}n
\]

\[
\widetilde{X}^{(n)}(x_{0},x)=n%
{\displaystyle\int\limits_{x_{0}}^{x}}
\widetilde{X}^{(n-1)}(x_{0},t)\frac{1}{t\varepsilon(t)}dt\text{ \ \ \ }%
\quad\text{for even }n
\]
\bigskip Then the formal powers in the meridional case are given by the
expressions \cite{Berskniga}%

\begin{align*}
_{\ast}Z^{(n)}(a^{\prime}+ia^{\prime\prime},z_{0};z)  &  =a^{\prime}%
{\displaystyle\sum\limits_{k=0}^{n}}
\binom{n}{k}X^{(n-k)}i^{k}(y-y_{0})^{k}\text{\ }\\
&  +ia^{\prime\prime}%
{\displaystyle\sum\limits_{k=0}^{n}}
\binom{n}{k}\widetilde{X}^{(n-k)}i^{k}(y-y_{0})^{k}\text{\ \ \ }\quad\text{for
odd }n
\end{align*}
and%

\begin{align*}
_{\ast}Z^{(n)}(a^{\prime}+ia^{\prime\prime},z_{0};z)  &  =a^{\prime}%
{\displaystyle\sum\limits_{k=0}^{n}}
\binom{n}{k}\widetilde{X}^{(n-k)}i^{k}(y-y_{0})^{k}\text{\ }\\
&  +ia^{\prime\prime}%
{\displaystyle\sum\limits_{k=0}^{n}}
\binom{n}{k}X^{(n-k)}i^{k}(y-y_{0})^{k}\text{\ \ \ \ }\quad\text{for even }n.
\end{align*}

\subsection{Formal powers in the transverse case}

As was shown in subsection 2.2 the Maxwell system (\ref{MaxwellStatic}) in the
transverse case reduces to the system%
\[
u_{x}=\frac{1}{\varepsilon}v_{y},\qquad u_{y}=-\frac{1}{\varepsilon}v_{x}%
\]
where $\varepsilon$ is a positive differentiable function of $r=\sqrt
{x^{2}+y^{2}}$. This system describing $\varepsilon$-analytic functions is
equivalent to the main Vekua equation (\ref{Vekuamain1}) where $f=\sqrt
{\varepsilon}$. In order to apply theorem \ref{ThGenSeq} we denote $u=\ln r$
and $U(u)=\sqrt{\varepsilon(e^{u})}$. Then taking $V\equiv1$ we obtain the
generating pair $(F,G)$ for equation (\ref{Vekuamain1}) in the desirable form%
\begin{equation}
F=U(u),\qquad G=\frac{i}{U(u)}. \label{GenpairPolar}%
\end{equation}
The analytic function $\Phi$ (from theorem \ref{ThGenSeq}) corresponding to
the polar coordinate system has the form $\Phi(z)=\ln z$ and consequently
$\Phi_{z}(z)=1/z$. We note that $\Phi_{z}$ has a pole in the origin and a zero
at infinity. Thus, theorem \ref{ThGenSeq} is applicable in any domain $\Omega$
which does not include these two points. Moreover, as for constructing formal
powers we need to use the recursive integration defined by (\ref{recformula})
in what follows we require $\Omega$ to be any bounded simply connected domain
not containing the origin.

From theorem \ref{ThGenSeq} we have that a generating sequence corresponding
to the generating pair (\ref{GenpairPolar}) can be defined as follows%
\[
F_{m}=\frac{U}{z^{m}}\quad\text{and}\quad G_{m}=\frac{i}{z^{m}U}\text{\quad
for even }m
\]
and
\[
F_{m}=\frac{1}{z^{m}U}\quad\text{and}\quad G_{m}=\frac{iU}{z^{m}}\text{\quad
for odd }m.
\]
As was explained in subsection \ref{SubsectMainVekua} in order to have a
complete system of formal powers for each $n$ we need to construct $Z_{{}%
}^{(n)}(1,z_{0};z)$ and $Z_{{}}^{(n)}(i,z_{0};z)$.

For $n=0$ we have
\[
Z_{{}}^{(0)}(1,z_{0};z)=\lambda_{1}^{(0)}F(z)+\mu_{1}^{(0)}G(z)
\]
and
\[
Z_{{}}^{(0)}(i,z_{0};z)=\lambda_{i}^{(0)}F(z)+\mu_{i}^{(0)}G(z)
\]
where $\lambda_{1}^{(0)}$, $\mu_{1}^{(0)}$ are real constants chosen so that
\[
\lambda_{1}^{(0)}F(z_{0})+\mu_{1}^{(0)}G(z_{0})=1
\]
and $\lambda_{i}^{(0)}$, $\mu_{i}^{(0)}$ are real constants such that
\[
\lambda_{i}^{(0)}F(z_{0})+\mu_{i}^{(0)}G(z_{0})=i.
\]
Taking into account that $F$ is real and $G$ is imaginary we obtain that
\[
\lambda_{1}^{(0)}=\frac{1}{F(z_{0})},\qquad\mu_{1}^{(0)}=0,
\]%
\[
\lambda_{i}^{(0)}=0,\qquad\mu_{i}^{(0)}=F(z_{0}).
\]
Thus,
\[
Z_{{}}^{(0)}(1,z_{0};z)=\frac{F(z)}{F(z_{0})}=\sqrt{\frac{\varepsilon
(r)}{\varepsilon(r_{0})}}%
\]
and
\[
Z_{{}}^{(0)}(i,z_{0};z)=\frac{iF(z_{0})}{F(z)}=i\sqrt{\frac{\varepsilon
(r_{0})}{\varepsilon(r)}}%
\]
where $r_{0}=\left\vert z_{0}\right\vert $.

For constructing $Z_{{}}^{(1)}(1,z_{0};z)$ and $Z_{{}}^{(1)}(i,z_{0};z)$ we
need first the formal powers $Z_{1}^{(0)}(1,z_{0};z)$ and $Z_{1}^{(0)}%
(i,z_{0};z)$. According to definition \ref{DefFormalPower} they have the form%
\[
Z_{1}^{(0)}(1,z_{0};z)=\lambda_{1}^{(1)}F_{1}(z)+\mu_{1}^{(1)}G_{1}(z)
\]
and
\[
Z_{1}^{(0)}(i,z_{0};z)=\lambda_{i}^{(1)}F_{1}(z)+\mu_{i}^{(1)}G_{1}(z)
\]
where $\lambda_{1}^{(1)}$, $\mu_{1}^{(1)}$ are real numbers such that
\[
\lambda_{1}^{(1)}F_{1}(z_{0})+\mu_{1}^{(1)}G_{1}(z_{0})=1
\]
and $\lambda_{i}^{(1)}$, $\mu_{i}^{(1)}$ are real numbers such that
\[
\lambda_{i}^{(1)}F_{1}(z_{0})+\mu_{i}^{(1)}G_{1}(z_{0})=i.
\]
Thus in order to determine $\lambda_{1}^{(1)}$, $\mu_{1}^{(1)}$ and
$\lambda_{i}^{(1)}$, $\mu_{i}^{(1)}$ we should solve two systems of linear
algebraic equations:%
\[
\lambda_{1}^{(1)}\frac{1}{z_{0}\varepsilon^{1/2}(r_{0})}+\mu_{1}^{(1)}%
\frac{i\varepsilon^{1/2}(r_{0})}{z_{0}}=1
\]
and%
\[
\lambda_{i}^{(1)}\frac{1}{z_{0}\varepsilon^{1/2}(r_{0})}+\mu_{i}^{(1)}%
\frac{i\varepsilon^{1/2}(r_{0})}{z_{0}}=i
\]
which can be rewritten as follows%
\[
\lambda_{1}^{(1)}+\mu_{1}^{(1)}i\varepsilon(r_{0})=\varepsilon^{1/2}%
(r_{0})z_{0}%
\]
and%
\[
\lambda_{i}^{(1)}+\mu_{i}^{(1)}i\varepsilon(r_{0})=i\varepsilon^{1/2}%
(r_{0})z_{0}.
\]
From here we obtain%
\[
\lambda_{1}^{(1)}=\varepsilon^{1/2}(r_{0})x_{0},\quad\mu_{1}^{(1)}%
=\varepsilon^{-1/2}(r_{0})y_{0},\quad\lambda_{i}^{(1)}=-\varepsilon
^{1/2}(r_{0})y_{0},\quad\mu_{i}^{(1)}=\varepsilon^{-1/2}(r_{0})x_{0}.
\]
Let us notice that in general for odd $m$ we have
\[
Z_{m}^{(0)}(1,z_{0};z)=\frac{\lambda_{1}^{(m)}}{z^{m}\varepsilon^{1/2}%
(r)}+\frac{i\mu_{1}^{(m)}\varepsilon^{1/2}(r)}{z^{m}},
\]
\[
Z_{m}^{(0)}(i,z_{0};z)=\frac{\lambda_{i}^{(m)}}{z^{m}\varepsilon^{1/2}%
(r)}+\frac{i\mu_{i}^{(m)}\varepsilon^{1/2}(r)}{z^{m}}%
\]
where
\[
\lambda_{1}^{(m)}=\varepsilon^{1/2}(r_{0})\operatorname*{Re}z_{0}%
^{m}=\varepsilon^{1/2}(r_{0})r_{0}^{m}\cos m\theta_{0},
\]%
\[
\mu_{1}^{(m)}=\varepsilon^{-1/2}(r_{0})\operatorname*{Im}z_{0}^{m}%
=\varepsilon^{-1/2}(r_{0})r_{0}^{m}\sin m\theta_{0},
\]
\[
\lambda_{i}^{(m)}=-\varepsilon^{1/2}(r_{0})\operatorname*{Im}z_{0}%
^{m}=-\varepsilon^{1/2}(r_{0})r_{0}^{m}\sin m\theta_{0},
\]%
\[
\mu_{i}^{(m)}=\varepsilon^{-1/2}(r_{0})\operatorname*{Re}z_{0}^{m}%
=\varepsilon^{-1/2}(r_{0})r_{0}^{m}\cos m\theta_{0},
\]
$\theta_{0}$ is the argument of the complex number $z_{0}$.

Thus, for odd $m$:%
\[
Z_{m}^{(0)}(1,z_{0};z)=\left(  \frac{r_{0}}{z}\right)  ^{m}\left(  \cos
m\theta_{0}\sqrt{\frac{\varepsilon(r_{0})}{\varepsilon(r)}}+i\sin m\theta
_{0}\sqrt{\frac{\varepsilon(r)}{\varepsilon(r_{0})}}\right)  ,
\]%
\[
Z_{m}^{(0)}(i,z_{0};z)=\left(  \frac{r_{0}}{z}\right)  ^{m}\left(  -\sin
m\theta_{0}\sqrt{\frac{\varepsilon(r_{0})}{\varepsilon(r)}}+i\cos m\theta
_{0}\sqrt{\frac{\varepsilon(r)}{\varepsilon(r_{0})}}\right)  .
\]
In a similar way we obtain the corresponding formulas for even $m$:%
\[
Z_{m}^{(0)}(1,z_{0};z)=\left(  \frac{r_{0}}{z}\right)  ^{m}\left(  \cos
m\theta_{0}\sqrt{\frac{\varepsilon(r)}{\varepsilon(r_{0})}}+i\sin m\theta
_{0}\sqrt{\frac{\varepsilon(r_{0})}{\varepsilon(r)}}\right)  ,
\]%
\[
Z_{m}^{(0)}(i,z_{0};z)=\left(  \frac{r_{0}}{z}\right)  ^{m}\left(  -\sin
m\theta_{0}\sqrt{\frac{\varepsilon(r)}{\varepsilon(r_{0})}}+i\cos m\theta
_{0}\sqrt{\frac{\varepsilon(r_{0})}{\varepsilon(r)}}\right)  .
\]
In order to apply formula (\ref{recformula}) for constructing formal powers of
higher orders we need to calculate the adjoint generating pairs $(F_{m}^{\ast
},G_{m}^{\ast})$. For odd $m$ we have
\[
F_{m}^{\ast}=-\frac{iz^{m}}{\varepsilon^{1/2}(r)},\quad G_{m}^{\ast
}=\varepsilon^{1/2}(r)z^{m}.
\]
For even $m$ we obtain
\[
F_{m}^{\ast}=-iz^{m}\varepsilon^{1/2}(r),\quad G_{m}^{\ast}=\frac{z^{m}%
}{\varepsilon^{1/2}(r)}.
\]
Now the whole procedure of construction of formal powers can be easily
algorithmized. The obtained system of formal powers
\[
\left\{  Z_{{}}^{(n)}(1,z_{0};z),\quad Z_{{}}^{(n)}(i,z_{0};z)\right\}
_{n=0}^{\infty}%
\]
is complete in the space of all solutions of the main Vekua equation
(\ref{Vekuamain1}) with $f=\varepsilon^{1/2}(r)$, i.e., any regular solution
$W$ of (\ref{Vekuamain1}) in $\Omega$ can be represented in the form of a
normally convergent series%
\[
W(z)=\sum_{n=0}^{\infty}Z_{{}}^{(n)}(a_{n},z_{0};z)=\sum_{n=0}^{\infty}\left(
a_{n}^{\prime}Z_{{}}^{(n)}(1,z_{0};z)+a_{n}^{\prime\prime}Z_{{}}^{(n)}%
(i,z_{0};z)\right)
\]
where $a_{n}^{\prime}=\operatorname*{Re}a_{n}$, $a_{n}^{\prime\prime
}=\operatorname*{Im}a_{n}$ and $z_{0}$ is an arbitrary fixed point in $\Omega$.

\end{document}